\RequirePackage{fix-cm}
\documentclass[twocolumn,epjc3]{svjour3}      
\RequirePackage{graphicx}
\RequirePackage{mathptmx}      
\RequirePackage{flushend}
\RequirePackage[numbers,sort&compress]{natbib}
\RequirePackage[colorlinks,citecolor=blue,urlcolor=blue,linkcolor=blue]{hyperref}
\RequirePackage[T1]{fontenc}
\journalname{Eur. Phys. J. C}
\smartqed  
\newcommand\beq{\begin{eqnarray}}
\newcommand\eeq{\end{eqnarray}}
\newcommand\bq{\begin{equation}}
\newcommand\eq{\end{equation}}
\newcommand\nutqpet{\mbox{\boldmath $(\hat{\eta}_{\nu})^{ \perp}$}\cdot({\bf \hat{q} \times
		(\hat{p}_{e})^\perp})}
\newcommand\netqpet{\mbox{\boldmath $(\hat{\eta}_{e})^{\perp}$}\cdot({\bf \hat{q} \times (\hat{p}_{e})^{\perp}})}
\newcommand\petnetnut{{\bf (\hat{p}_{e})^{\perp}}\cdot( \mbox{\boldmath $(\hat{\eta}_{e})^{\perp}$}\times \mbox{\boldmath $(\hat{\eta}_{\nu})^{ \perp}$})}
\newcommand\qnetnut{{\bf \hat{q}}\cdot( \mbox{\boldmath $(\hat{\eta}_{e})^{\perp}$}\times \mbox{\boldmath $(\hat{\eta}_{\nu})^{ \perp}$})}
\newcommand\netnut{\mbox{\boldmath $(\hat{\eta}_{e})^{\perp}$}\cdot\mbox{\boldmath $(\hat{\eta}_{\nu})^{ \perp}$}}
\newcommand\nutpet{\mbox{\boldmath $(\hat{\eta}_{\nu})^{ \perp}$}\cdot{\bf (\hat{p}_{e})^{\perp}}}

\newcommand\netpet{\mbox{\boldmath $(\hat{\eta}_{e})^{\perp}$}\cdot{\bf (\hat{p}_{e})^{\perp}}}
\DeclareSymbolFont{matha}{OML}{txmi}{m}{it}
\DeclareMathSymbol{\vv}{\mathord}{matha}{29} 

\begin{document}
	\title{Neutrino elastic scattering on polarized electrons  as a tool for probing the neutrino nature} 
	\author{A. B\l{}aut\thanksref{e2,addr1} \and W. Sobk\'ow\thanksref{e1,addr1}  
	}
	\thankstext{e2}{e-mail: arkadiusz.blaut@uwr.edu.pl}
	\thankstext{e1}{e-mail: wieslaw.sobkow@uwr.edu.pl (corresponding author)}
	
	
	\institute{ Institute of Theoretical Physics, University of Wroc\l{}aw,
		Pl. M. Born 9, PL-50-204~Wroc{\l}aw, Poland\label{addr1}
	}
	
	\date{Received: date / Accepted: date}
	
	\maketitle
\begin{abstract}

Possibility of using the polarized electron target (PET) for testing the neutrino nature is considered.
 It is assumed that the incoming $\nu_e$ beam is the  superposition of left chiral (LC) states with right chiral (RC) ones. Consequently the non--vanishing transversal components of $\nu_e$ spin polarization may appear, both T-even and T-odd. $\nu_e$s are produced by the low energy monochromatic  (un)polarized emitter located at a  near distance from  the hypothetical detector which is   able to measure  both the azimuthal angle  and the polar angle  of the recoil electrons, and/or also the energy of the outgoing electrons with a high resolution.    A detection process is the elastic scattering of $\nu_e$s  (Dirac or Majorana) on the polarized electrons. LC $\nu_e$s interact mainly by the standard  $V - A$ interaction, while RC ones participate only  in the non--standard $V + A$, scalar $S_R$,  pseudoscalar $P_R$ and  tensor $T_R$ interactions. All the interactions are of  flavour-conserving type (FC).
We show that  a distinction between the Dirac and the  Majorana $\nu_e$s is possible both for the longitudinal and the transversal $\nu_e$ polarizations.
\\
In the first case a departure from the standard  prediction of the azimuthal asymmetry of recoil electrons  is caused by the interferences between the non-standard complex S and T couplings, proportional to the  angular correlations (T-even and T-odd) among the polarization of
the electron target, the incoming neutrino momentum and the outgoing electron momentum. 
It is shown that such a deviation would indicate the Dirac $\nu_e$ nature and the presence of time reversal symmetry violation (TRSV) interactions. 
It is remarkable that the result is conclusive for all Majorana non--standard couplings.
\\
In the second case the azimuthal asymmetries, polar distribution and energy spectrum of scattered electrons are sensitive to the interference terms between the standard and exotic interactions, proportional to the various  angular correlations among the transversal
$\nu_e$  spin polarization,
the electron target polarization, the incoming $\nu_e$ momentum and the outgoing electron momentum.
In the particular case of the $V-A$ and $S$ couplings the precise measurement of some observables, e.g. the spectrum,
can distinguish between the Dirac and the Majorana $\nu_e$s as long as the incoming $\nu_e$ beam has non-vanishing transversal polarization. Our model-in\-de\-pen\-dent study is carried out for the flavour $\nu_e$ eigenstates in the  relativistic $\nu_e$ limit.

\end{abstract}

\section{Introduction}
\label{sec1}
One of the basic questions in  neutrino physics is whether the $\nu$s  are Dirac or Majorana fermions. At present, the neutrinoless double beta decay is viewed as the main tool to investigate  $\nu$s nature \cite{Majorana,Majorana1,Majorana2}, however the purely leptonic processes  (e.g. the neutrino-electron elastic scattering (NEES)) may also shed some light on this problem \cite{Kayser,Langacker}. Kayser and Langacker have analyzed the $\nu$s nature problem in the context of non--zero $\nu$s mass and of the  standard model (SM) V-A interaction \cite{SM,SM1,SM2,SM3,SM4} of only the LC $\nu$s. There is  an alternative opportunity  of distinguishing between Majorana and Dirac $\nu$s by admitting the exotic  $V + A$, scalar $S$,  pseudoscalar $P$ and  tensor $T$ interactions coupling to the LC and  RC  $\nu$s in the leptonic processes within the relativistic  $\nu$ limit. The appropriate tests have been considered by Rosen \cite{Rosen} and Dass \cite{Dass} (see also \cite{Zralek,Zralek1,Nishiura,Semikoz,Pastor,Singh,Gutierrez,RXY,Xu,Gouvea,Gouvea1,Grimus} for other works devoted to the  $\nu$ nature and the non--standard $\nu$ properties). The above ideas involve the unpolarized detection target. When the target-electrons are polarized by an external magnetic field, one has a possibility of changing the rate of weak interaction by inverting the direction of magnetic field. This feature is very important in the detection of low energy $\nu_e$s because the background level would be precisely controlled \cite{Misiaszek}. PET seems to be  a more sensitive laboratory for probing the  $\nu$ nature and  TSRV in the leptonic processes than the unpolarized target due to the mentioned control of contribution of the interaction to the cross section.  It is worth reminding that the PET  has been proposed to test the flavour composition of  (anti)neutrino beam \cite{PET3} and various effects of non--standard physics. We mean the neutrino magnetic moments  \cite{PET4,PET}, TRSV in the (semi)leptonic processes  \cite{PET2,SBPET}, axions, spin--spin interaction in gravitation   \cite{PET1,PET5,PET6,PET7}. The possibility of using polarized targets of nucleons and of  electrons for the fermionic, scalar and   vector dark matter  detection is also worth noticing \cite{DM1,DM2,DM3}. The methods  of  producing  the  spin-polarized  gasses such as helium,  argon and xenon  are described in \cite{Gass1,Gass2}.\\
It is also essential to mention the  measurements confirming  the possibility of realizing the polarized target crystal of $Gd_{2}SiO_{5}$ (GSO) doped with Cerium (GSO:Ce) \cite{INFN}. \\ Let us recall that there is no difference between Dirac and Majorana $\nu$s in the case of NEES with the  standard  V-A interaction in the relativistic limit, when the target is unpolarized.  
\\
The SM does not allow the clarification of the origin of  parity violation,  observed  baryon asymmetry of universe \cite{barion} through a single CP-violating phase of the Cabibbo-Ko\-ba\-ya\-shi-Mas\-ka\-wa quark-mixing matrix (CKM) \cite{Kobayashi}  and other fundamental problems. This has led to the appearance of many non-standard models: the left-right symmetric models (LRSM) \cite{Pati,Pati1,Pati2,Pati3}, composite models \cite{Jodidio,CM,CM1}, models with extra dimensions (MED) \cite{Extra} and  the unparticle  models (UP) \cite{unparticle,unparticle1,unparticle2}. There is a rich literature devoted to the phenomenological  aspects of the non--standard interactions  of LC and in particular RC $\nu$s: \cite{unparticle3,unparticle5,unparticle6,unparticle7,unparticle8,unparticle9,unparticle10,unparticle12,NSI,NSI1,NSI2,NSI3,massivenu,massivenu1,massivenu2,massivenu3,massivenu4,fenNSI,fenNSI1,fenNSI4,
fenNSI5,fenNSI6,fenNSI7,fenNSI8,fenNSI9,fenNSI10,fenNSI11,
fenNSI14}. It is also  noteworthy  that the current experimental results still leave some space for the scenarios with the exotic interactions.  \\ Recently the study  of the $\nu$ nature  with a use of PET in the case of  standard V-A interaction, when the evolution of $\nu$ spin polarization in the astrophysical environments is admitted,  has been carried out in \cite{Barranco,Barranco1}. \\
In this paper we consider   the elastic  scattering of  low energy $\nu_e$s ($\sim 1 MeV$)  on the polarized electrons of target in the presence of non-standard complex scalar, pseudoscalar, tensor couplings and $V+A$ interaction as a useful tool for testing the $\nu$ nature. We show how the various types of azimuthal asymmetry, the polar distribution and the energy spectrum of scattered electrons enable the  distinguishing  between the Dirac and the  Majorana $\nu_e$s  both for the  longitudinal and the transversal $\nu_e$ polarizations,
 taking into account TRSV. Both theoretically possible  scenarios of physics beyond  SM deal with  FC standard and non--standard interactions.
Our study is model--independent and carried out for the flavour $\nu_e$ eigenstates  (Dirac and Majorana)  in the relativistic limit. One assumes that  the mo\-no\-chro\-matic  low energy  and (un)po\-la\-rized  $\nu_e$  emitter with  a high activity is placed at a  near distance from the detector (or at the detector centre). 
The hypothetical detector is assumed to be able to measure  both the azimuthal angle $\phi_{e}$ and the polar angle $\theta_e$ of the recoil  electrons, and/or also the energy of the outgoing electrons with a high resolution, Fig.1.  We  utilize the experimental values of standard couplings: $c_{V}^{L}= 1 -0.04, c_{A}^{L}= 1-0.507$ to  evaluate the predicted  effects \cite{Data}, where the indexes $V, A$ and $L$ denote the vector, the axial interactions and the left-handed chirality of $\nu_e$s, respectively. We assume the values of exotic couplings which are compatible with the constraints on non--standard interactions obtained in our previous paper \cite{SBSOX}  and  by various authors: \cite{Deniz1,Deniz2}.
The laboratory differential cross sections (see Appendix 1 for Majorana $\nu_e$s and \cite{SBPET} for the Dirac case) are calculated with   the use of the covariant
projectors   for the incoming $\nu_{e}$s (including both the longitudinal and the transversal components of the spin polarization) in the relativistic limit   and  for the polarized target-electrons, respectively \cite{Michel}.
\section{Elastic scattering of  Dirac  electron neutrinos  on polarized electrons}
\label{sec2}
\begin{figure}
    \begin{center}
        \includegraphics[width=1\linewidth]{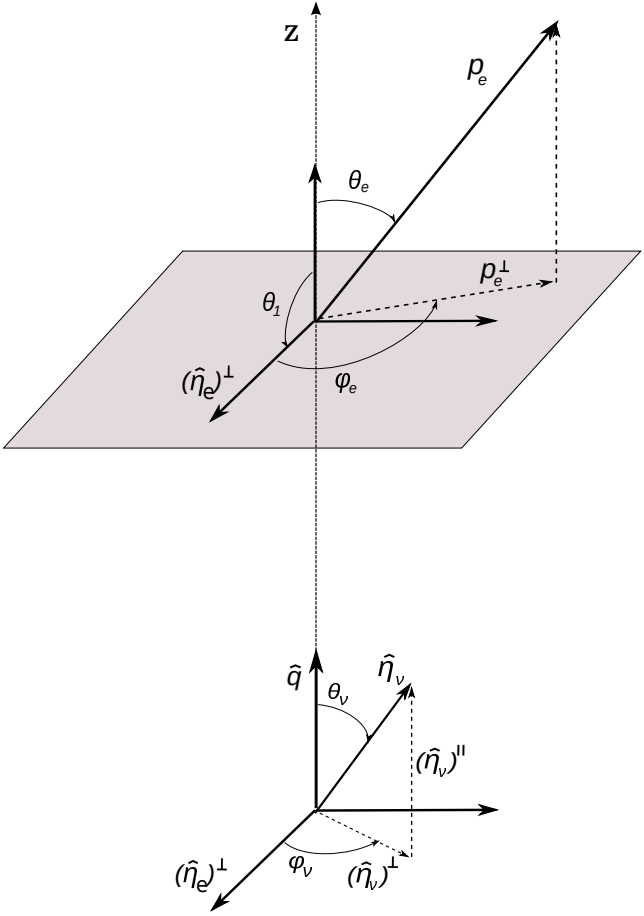}
    \end{center}
    \caption{
        The reaction  plane is spanned by the $\nu_{e}$  LAB momentum unit vector $\hat{\bf q}$ and the electron
        polarization vector of the target $\mbox{\boldmath $ \hat{\eta}_{e} $}$.  $\theta_1$ is the angle  between $\mbox{\boldmath $\hat{\eta}_{e}$}$ and  $\hat{\bf q}$ (on the plot $\theta_1=\pi/2$ so in this case $\mbox{\boldmath $\hat{\eta}_{e}$}=\mbox{$(\boldmath \hat{\eta}_{e})^\perp $}$).
        $\theta_{e}$ is the polar angle between ${\bf \hat{q}}$ and
        the unit vector $ \hat{\bf p}_{e}$ of the recoil electron momentum. $\phi_{e}$ is the
        angle between $\mbox{(\boldmath $ \hat{\eta}_{e})^\perp $}$ and the transversal component of  outgoing electron momentum $ (\hat{\bf p}_{e})^{\perp}$.
        $\phi_{\nu}$ and $\theta_{\nu}$ are the azimuthal and the polar angles of the unit polarization vector of the incoming neutrino, 
        $\mbox{\boldmath $\hat{\eta}_{\nu}$}=(\sin\theta_\nu \,\cos\phi_{\nu}, \sin\theta_\nu \,\sin\phi_{\nu}, \cos\theta_\nu)$.
        \label{Fig1D}}
\end{figure}
We  analyze a scenario in which    the incoming  Dirac $\nu_{e}$ beam  is assumed to be the superposition of LC states with RC ones. The detection process is the elastic scattering of  Dirac $\nu_{e}$s on the polarized target-electrons. 
The relative orientation of the incoming $\nu_{e}$, the polarized electron target and the outgoing electron is depicted in  Fig. 1, where the notation for the relevant quantities is explained. 
\\
LC $\nu_e$s interact mainly by the standard $V - A$ interaction and a small admixture of the non-standard  scalar $S_L$, pseudoscalar $P_L$, tensor $T_L$ interactions, while  RC ones take part only in the exotic $V + A$ and $S_R, P_R, T_R$ interactions. As a result of the superposition of the two chiralities the  spin  polarization vector has the non-vanishing transversal polarization components, which may give rise to both
      T-even and T-odd effects. As an example of the process in which the transversal $\nu$ polarization may be produced, we refer to the ref. \cite{CMS2003},  where the muon capture by proton  has been considered.
The amplitude for the   $\nu_{e}  e^{-}$ scattering in  low energy region is in the form:
\beq \label{ampD} M^{D}_{\nu_{e} e^{-⁠}}
&=&
\frac{G_{F}}{\sqrt{2}}\{(\overline{u}_{e'}\gamma^{\alpha}(c_{V}^{L}
-⁠ c_{A}^{L}\gamma_{5})u_{e}) (\overline{u}_{\nu_{e'}}
\gamma_{\alpha}(1 -⁠ \gamma_{5})u_{\nu_{e}})\nonumber\\
&& \mbox{} + (\overline{u}_{e'}\gamma^{\alpha}(c_{V}^{R}
+ c_{A}^{R}\gamma_{5})u_{e}) (\overline{u}_{\nu_{e'}}
\gamma_{\alpha}(1 + \gamma_{5})u_{\nu_{e}})  \\
&  & \mbox{} +
c_{S}^{R}(\overline{u}_{e'}u_{e})(\overline{u}_{\nu_{e'}}
(1 + \gamma_{5})u_{\nu_{e}}) \nonumber\\
&  & \mbox{} +
c_{P}^{R}(\overline{u}_{e'}\gamma_{5}u_{e})(\overline{u}_{\nu_{e'}}
\gamma_{5}(1 + \gamma_{5})u_{\nu_{e}}) \nonumber\\
&& \mbox{} +
\frac{1}{2}c_{T}^{R}(\overline{u}_{e'}\sigma^{\alpha \beta}u_{e})(\overline{u}_{\nu_{e'}}
\sigma_{\alpha \beta}(1 + \gamma_{5})u_{\nu_{e}})\nonumber\\
&&\mbox{} +
c_{S}^{L}(\overline{u}_{e'}u_{e})(\overline{u}_{\nu_{e'}}
(1 -⁠ \gamma_{5})u_{\nu_{e}}) \nonumber\\
&&\mbox{} + c_{P}^{L}(\overline{u}_{e'}\gamma_{5}u_{e})(\overline{u}_{\nu_{e'}}
\gamma_{5}(1 -⁠ \gamma_{5})u_{\nu_{e}}) \nonumber\\
&& \mbox{} +
\frac{1}{2}c_{T}^{L}(\overline{u}_{e'}\sigma^{\alpha \beta}u_{e})(\overline{u}_{\nu_{e'}}
\sigma_{\alpha \beta}(1 -⁠ \gamma_{5})u_{\nu_{e}})
\}\nonumber,
\eeq
where $G_{F} = 1.1663788(7)\times
10^{-5}\,\mbox{GeV}^{-2} (0.6 \; ppm)$ \cite{Mulan} is the Fermi constant. The coupling constants are
denoted as $c_{V}^{L, R} $,
$c_{A}^{L, R}$, $c_{S}^{R, L}$, $c_{P}^{R, L}$,  $c_{T}^{R, L}$ respectively to the incoming $\nu_{e}$
of left- and right-handed chirality. All  the non-standard couplings  $c_{S}^{R, L}$, $c_{P}^{R, L}$,  $c_{T}^{R, L}$  are the complex numbers denoted as
$c_S^R = |c_S^R|e^{i\,\theta_{S,R}}$, $c_S^L = |c_S^L|e^{i\,\theta_{S,L}}$, etc. Reality of  $c_{V}^{L, R} $,
$c_{A}^{L, R}$ coupling  constants follows from the  hermiticity of the interaction lagrangian; for the same reason we take into account the relations between the non-standard complex couplings with left- and right-handed chirality,
 $c_{S, T, P}^{ L}=c_{S, T, P}^{*R}$. All results are stated in terms of $R$ couplings. 
\section{Elastic scattering of  Majorana electron neutrinos  on polarized electrons}
\label{sec3}
The fundamental difference between Majorana and Dirac  $\nu_e$s arises from a fact that Majorana $\nu_e$s do not participate in the vector V and tensor T interactions.  This is a direct consequence of the $(u,\vv)$--mode decomposition of the Majorana field.
The amplitude for NEES on PET  for Majorana  low energy $\nu_e$s is as follows:
\beq \label{ampM} \lefteqn{ M^{M}_{\nu_{e} e^{-⁠}}
    =  \mbox{} \frac{2G_{F}}{\sqrt{2}}\{-(\overline{u}_{e'}\gamma^{\alpha}(c_{V}
    - c_{A}\gamma_{5})u_{e}) (\overline{u}_{\nu_{e'}}\gamma_{\alpha}\gamma_{5} u_{\nu_{e}}) }\\
& + & (\overline{u}_{e'}u_{e})\left[c_{S}^{L}(\overline{u}_{\nu_{e'}}
(1 - \gamma_{5})u_{\nu_{e}}) +  c_{S}^{R}(\overline{u}_{\nu_{e'}}(1 + \gamma_{5})u_{\nu_{e}})\right]\nonumber \\
& + &  (\overline{u}_{e'}\gamma_{5}u_{e})\left[ -c_{P}^{L}(\overline{u}_{\nu_{e'}}
(1 - \gamma_{5})u_{\nu_{e}}) +  c_{P}^{R}(\overline{u}_{\nu_{e'}}(1 + \gamma_{5})u_{\nu_{e}})\right]\}.  \nonumber
\eeq
 We see that the $\nu_{e}$   contributions from $A, S, P$  are multiplied by the factor of $2$ as a result of the  Majorana condition.
 The indexes $L$, ($R$) for the standard interactions  are omitted. It means that both LC and RC $\nu_{e}$s may take part in the above  interactions.
  All the other assumptions are the same as for the Dirac case.
\section{ Distinguishing between Dirac and Majorana neutrinos through azimuthal asymmetries of recoil electrons}
\label{sec4}
\begin{figure}
    \begin{center}
        \includegraphics[width=0.9\linewidth]
        {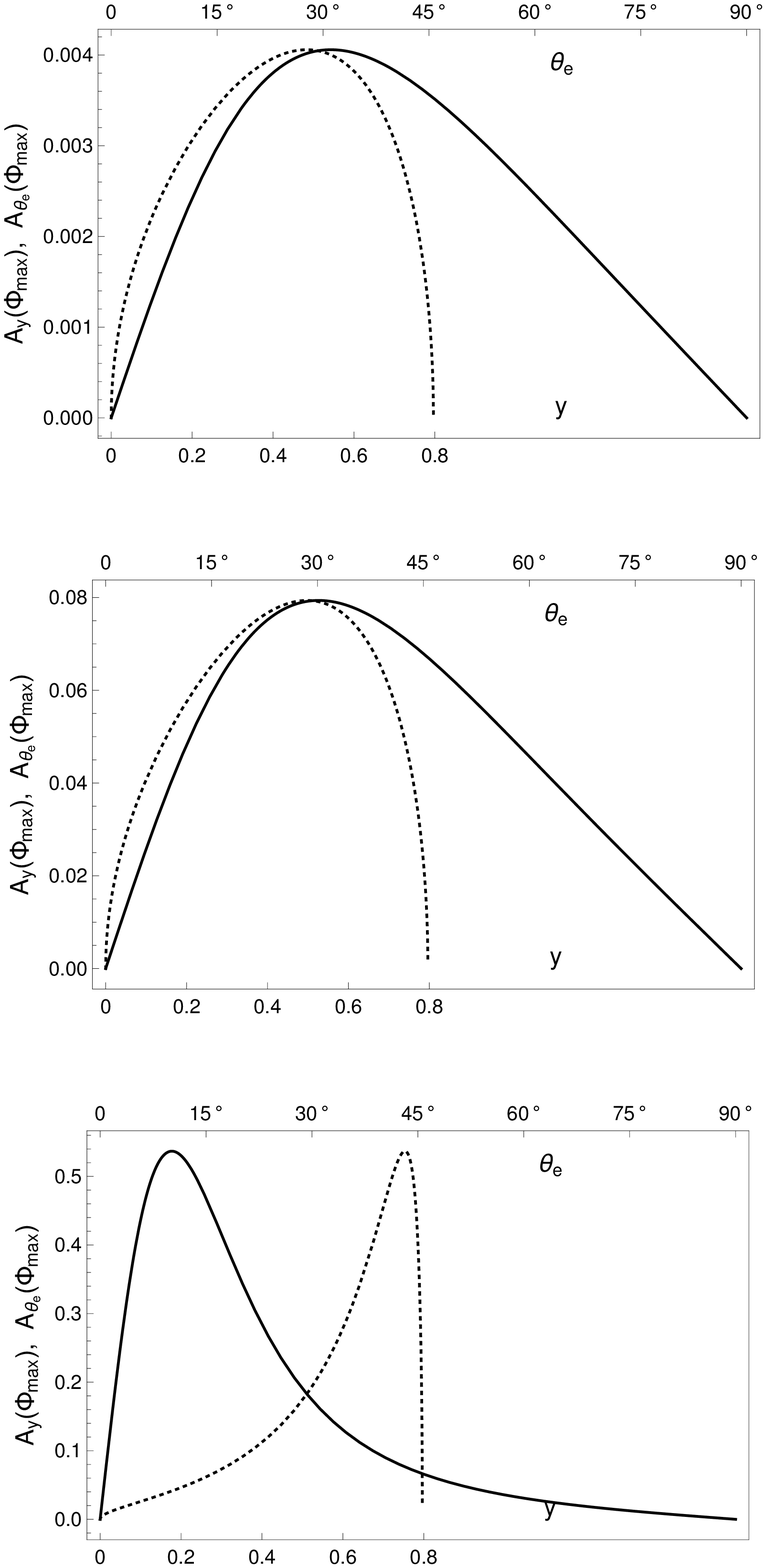}
        \caption{Dirac (or Majorana) $\nu_e$ with    $V-A$ interaction: plot of     $A_y(\Phi_{max})$ as a function y (dotted line) and $A_{\theta_e}(\Phi_{max})$ as a function of $\theta_e$ (solid line) for $\mbox{\boldmath $\hat{\eta}_{\nu}$}\cdot\hat{\bf q}=-1$, $E_\nu=1\,MeV$, $\Phi_{max}=\pi/2$: upper plot for $\theta_1=0.1$; middle plot for $\theta_1=\pi/2$; lower plot for $\theta_1=\pi-0.1$. \label{Fig.2}}
    \end{center}
\end{figure}
\begin{figure}
    \begin{center}
        \includegraphics[width=0.95\linewidth]
{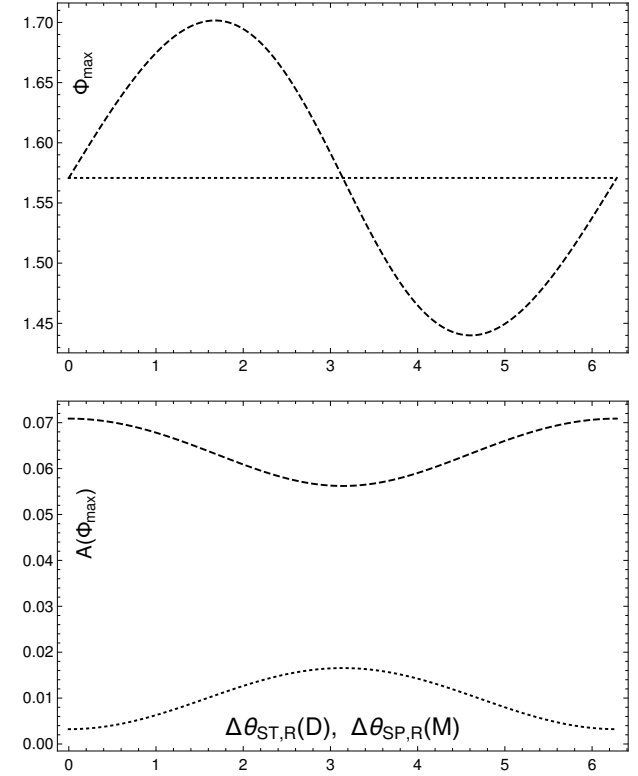}
        \caption{Upper plot (dashed line) is  the  dependence of $\Phi_{max}$    on 
        	$\Delta \theta_{ST,R}(D)$ for 	Dirac $\nu_e$ in case  of  $V-A$, $S_R$ and $T_R$ interactions when $ |c_S^R|=|c_T^R|=0.2$.  Upper plot (dotted line) is  the  dependence of $\Phi_{max}$ on $\Delta \theta_{SP,R}(M)$ for	Majorana $\nu_e$ in case  of  $V-A$, $S_R$ and $P_R$ couplings when $|c_S^R|=|c_P^R|=0.2$. Both scenarios assume $\mbox{\boldmath $\hat{\eta}_{\nu}$}\cdot\hat{\bf q}=-1$, 
        	$ \theta_1=\pi/2, E_\nu=1\,MeV$.  Lower plot is the dependence of  $A(\Phi_{max})$  on $\Delta \theta_{ST,R}(D)$ for 	Dirac $\nu_e$ (dashed line) and  on $\Delta \theta_{SP,R}(M)$ for 	Majorana $\nu_e$ (dotted line), respectively, with  same assumptions as for $\Phi_{max}$. 
        	\label{Fig.5} }
    \end{center}
\end{figure}
\begin{figure}
    \begin{center}
        \includegraphics[width=0.98\linewidth]
        {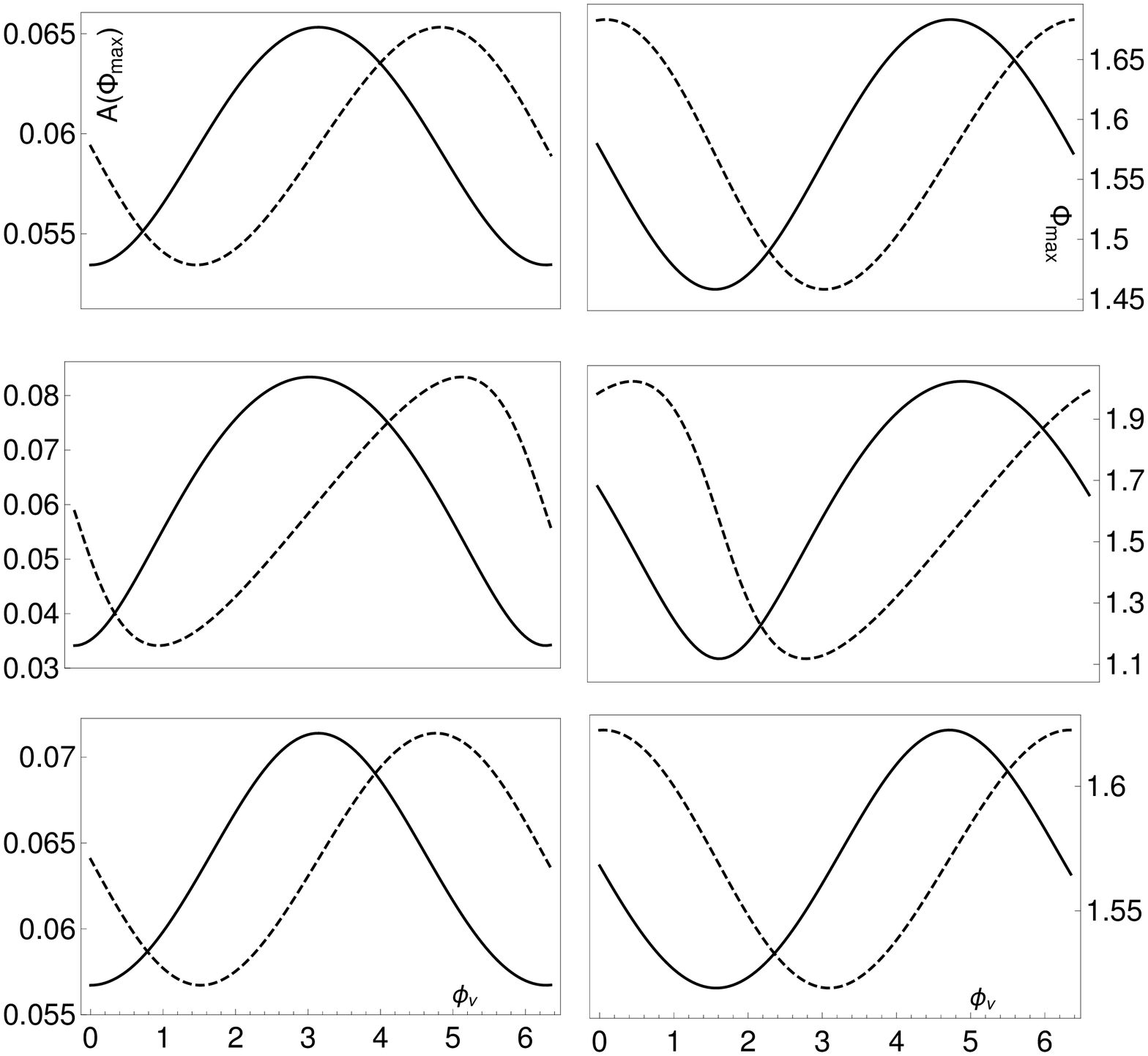}
        \caption{Superposition of LC $\nu_e$s with RC ones in presence of non--standard couplings with $\mbox{\boldmath $\hat{\eta}_{\nu}$}\cdot\hat{\bf q}=-0.95$: dependence of $A(\Phi_{max})$ on $\phi_\nu$  (solid line) and $\Phi_{max}$ on $\phi_\nu$ (dashed line) for $ E_\nu=1\,MeV, \theta_1=\pi/2$. TRSC: upper left plot for  Dirac case of   $V-A $ and $S_R$ when $|c_S^R|=0.2, \theta_{S,R}=0$; middle left plot for Majorana case of $V-A $ with $S_R$ when $|c_S^R|=0.2, \theta_{S,R}=0$; lower left plot for  Dirac case of $V-A $ with $T_R$ when $|c_T^R|=0.2, \theta_{T,R}=0$. TRSV:  upper right plot for Dirac  scenario with $V-A $ and $S_R$ when $|c_S^R|=0.2, \theta_{S,R}=\pi/2$;  middle right plot for Majorana case   of $V-A $ with $S_R$ when $|c_S^R|=0.2, \theta_{S,R}=\pi/2$; lower right plot for Dirac case of $V-A $ and $T_R$ when $|c_T^R|=0.2, \theta_{T,R}=\pi/2$. \label{Fig.7} }
    \end{center}
\end{figure}
\begin{figure}
    \begin{center}
        \includegraphics[width=0.99\linewidth]
{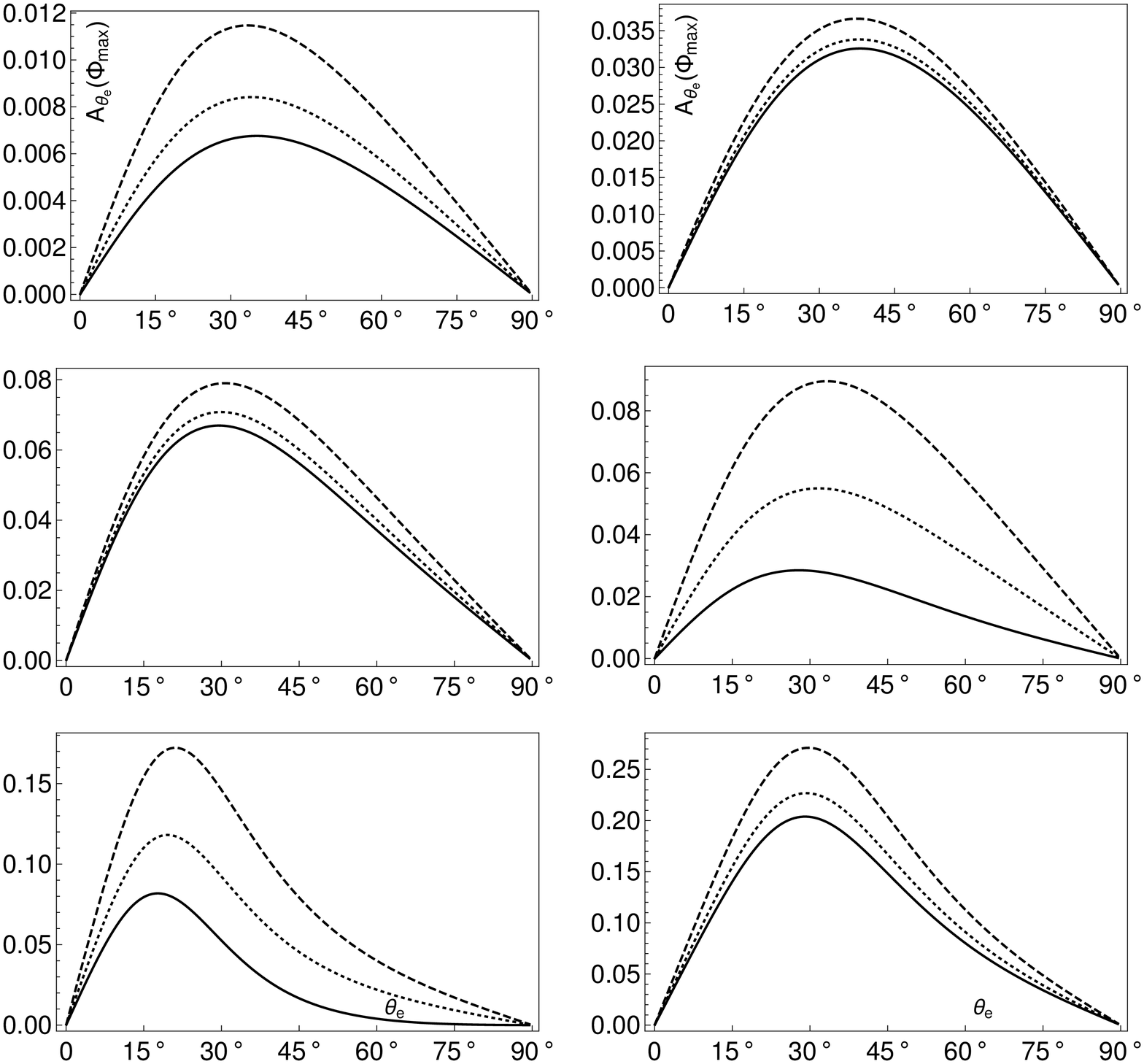}
        \caption{Superposition of LC $\nu_e$s with RC ones in presence of non--standard couplings with $\mbox{\boldmath $\hat{\eta}_{\nu}$}\cdot\hat{\bf q}=-0.95$:    plot of  $A_{\theta_e}(\Phi_{max})$ as a function of $\theta_e$  for the case of $V-A$ with $S_R$ when   $E_\nu=1\,MeV$, $\phi_\nu=0$;left column for Dirac $\nu_e$, right column for Majorana $\nu_e$; upper plot for $\theta_1=0.1$; middle plot for $\theta_1=\pi/2$; lower plot for $\theta_1=\pi-0.1$; solid line for $|c_S^R|=0.3$, $\theta_{S,R}=0$; dotted line for  $|c_S^R|=0.3$, $\theta_{S,R}=\pi/4$; dashed line for  $|c_S^R|=0.3$, $\theta_{S,R}=\pi/2$. \label{Fig.8}}
    \end{center}
\end{figure}
\begin{figure}
    \begin{center}
        \includegraphics[width=0.99\linewidth]
{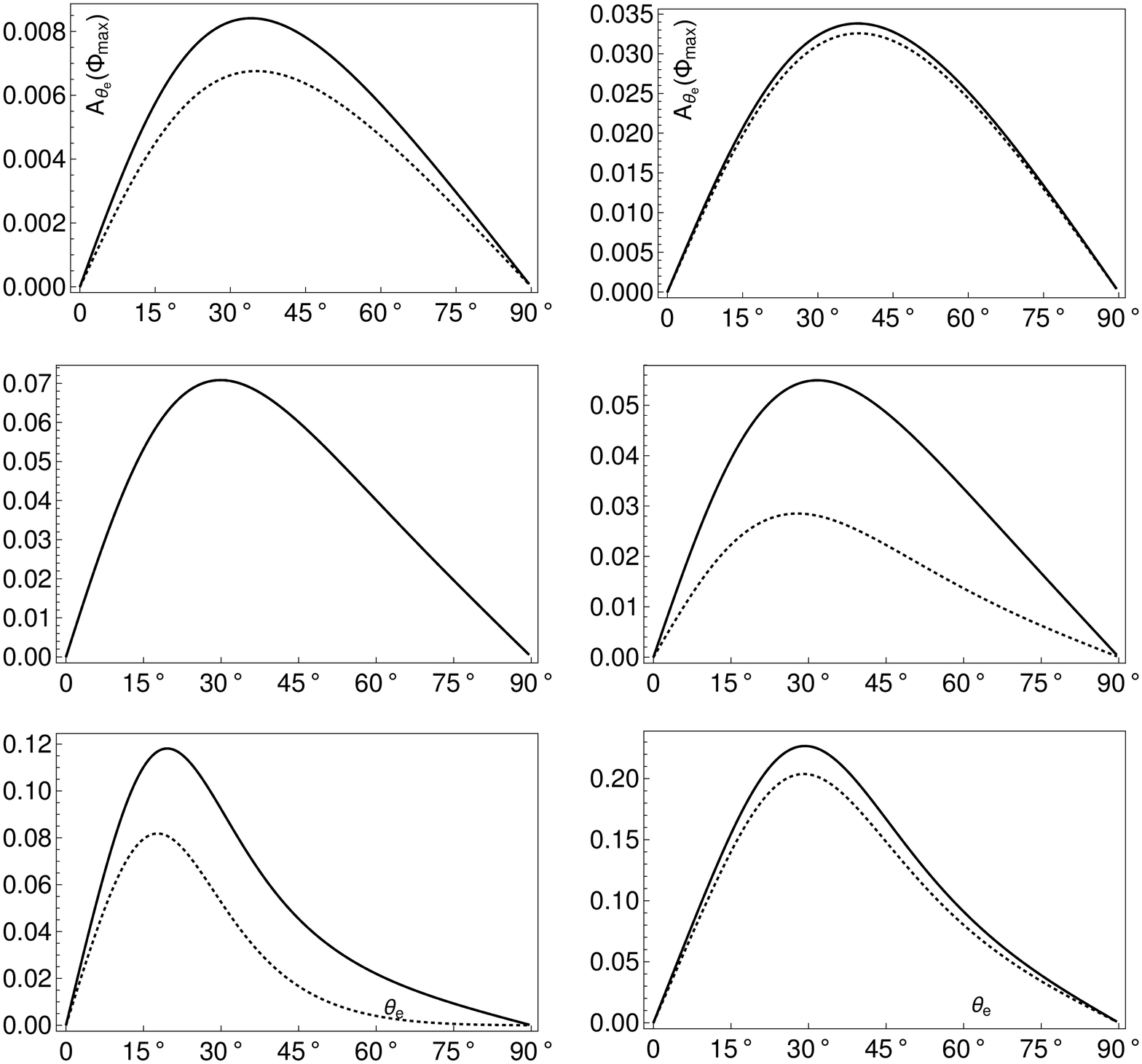}
        \caption{Superposition of LC $\nu_e$s with RC ones in presence of non--standard couplings with $\mbox{\boldmath $\hat{\eta}_{\nu}$}\cdot\hat{\bf q}=-0.95$:    plot of  $A_{\theta_e}(\Phi_{max})$ as a function of $\theta_e$  for the case of $V-A$ with $S_R$ when   $E_\nu=1\,MeV$, $\phi_\nu=\pi/4$;left column for Dirac $\nu_e$, right column for Majorana $\nu_e$; upper plot for $\theta_1=0.1$; middle plot for $\theta_1=\pi/2$; lower plot for $\theta_1=\pi-0.1$; solid line for $|c_S^R|=0.3$, $\theta_{S,R}=0$; dotted line for  $|c_S^R|=0.3$, $\theta_{S,R}=\pi/4$;dashed line for  $|c_S^R|=0.3$, $\theta_{S,R}=\pi/2$. \label{Fig.9}}
    \end{center}
\end{figure}
In this section we analyze the possibility of distinguishing the  Dirac from the  Majorana $\nu_{e}$s through probing the  azimuthal asymmetries, $A$, $A_y$, $A_{\theta_e}$, of recoil electrons. The asymmetry functions are defined by the following formulas:
\beq
A(\Phi) := \frac{\int\limits_{\Phi}^{\Phi+\pi}\frac{d\sigma}{d \phi_{e}}\,d\phi_e -
    \int\limits_{\Phi+\pi}^{\Phi+2\pi}\frac{d\sigma}{d \phi_{e}}\,d\phi_e}
{\int\limits_{\Phi}^{\Phi+\pi}\frac{d\sigma}{d \phi_{e}}\,d\phi_e +
    \int\limits_{\Phi+\pi}^{\Phi+2\pi}\frac{d\sigma}{d \phi_{e}}\,d\phi_e},
\eeq
\beq
A_y(\Phi) := \frac{\int\limits_{\Phi}^{\Phi+\pi}\frac{d^2\sigma}{d \phi_{e} dy}\,d\phi_e -
    \int\limits_{\Phi+\pi}^{\Phi+2\pi}\frac{d^2\sigma}{d \phi_{e} dy}\,d\phi_e}
{\int\limits_{\Phi}^{\Phi+\pi}\frac{d^2\sigma}{d \phi_{e} dy}\,d\phi_e +
    \int\limits_{\Phi+\pi}^{\Phi+2\pi}\frac{d^2\sigma}{d \phi_{e} dy}\,d\phi_e},
\eeq
\beq
A_{\theta_e}(\Phi) := \frac{\int\limits_{\Phi}^{\Phi+\pi}\frac{d^2\sigma}{d \phi_{e}d\theta_e}\,d\phi_e -
    \int\limits_{\Phi+\pi}^{\Phi+2\pi}\frac{d^2\sigma}{d \phi_{e} d\theta_e}\,d\phi_e}
{\int\limits_{\Phi}^{\Phi+\pi}\frac{d^2\sigma}{d \phi_{e} d\theta_e}\,d\phi_e +
    \int\limits_{\Phi+\pi}^{\Phi+2\pi}\frac{d^2\sigma}{d \phi_{e} d\theta_e}\,d\phi_e}.
\eeq
These observables are functions of the asymmetry angle $\Phi$ (the interpretation of $\Phi$ follows from the definitions  of eqs. (3-5)  and  Fig. 1: $\Phi$ is measured with respect to  the  transverse electron
polarization vector of target $\mbox{(\boldmath $ \hat{\eta}_{e})^\perp $}$); by $\Phi_{max}$ we denote the location of their maxima.
In what follows we shall use the normalized, dimensionless kinetic energy of the recoil electrons $y$, defined by 
\beq y & \equiv &
\frac{T_e}{E_\nu}=\frac{m_{e}}{E_{\nu}}\frac{2 \cos^2(\theta_{e})}{(1+\frac{m_{e}}{E_{\nu}})^{2}-\cos^2(\theta_{e})},
\eeq
where $T_{e}$  is the
kinetic energy of the recoil electron, $E_{\nu}$   is the incoming $\nu_e$ energy, $m_{e}$ is the electron mass. 
 Fig. 2  shows the asymmetries  $A_y(\Phi_{max})$,  $A_{\theta_e}(\Phi_{max})$ for the standard  V-A interaction with  
$\mbox{\boldmath $\hat{\eta}_{\nu}$}\cdot\hat{\bf q}=-1$. 
Although the orientation of the asymmetry axis is fixed at $\Phi_{max}=\pi/2$ the magnitude of the asymmetries may change with $y$ and $\theta_e$.
We see that the maximum values of $A_y(\Phi_{max})$ and  $A_{\theta_e}(\Phi_{max})$
also depend on the angle $\theta_1$ between $\mbox{\boldmath $\hat{\eta}_{e}$}$ and  $\hat{\bf q}$: they
grow from  $0.004$ for $\theta_1=0.1$ (upper plot) to $0.54$ for $\theta_1=\pi-0.1$ (lower plot). 
However all curves on the diagrams are identical for the Dirac and the  Majorana $\nu_{e}$s therefore 
the asymmetries can not discriminate between the two $\nu$ types even if the target-electrons are polarized. 

The presence of  non--stan\-dard $S, T, P$ complex couplings of  Dirac $\nu_e$s with $\mbox{\boldmath $\hat{\eta}_{\nu}$}\cdot\hat{\bf q}=-1$ leads to the non--va\-ni\-shing triple angular correlations composed of $ \hat{\bf q}, \hat{\bf p}_{e}, \mbox{\boldmath $(\hat{\eta}_{e})^{\perp}$}$ vectors. These terms not only play a role in distinguishing between the Dirac and the  Majorana $\nu_{e}$s but in the Dirac case they allow 
to search for the effects of TRSV in NEES.   Fig. 3 shows how the  asymmetry axis location $\Phi_{max}$ (upper plot) and the magnitude of $A(\Phi_{max})$ (lower plot) depend on the phase differences $\Delta \theta_{ST,R}(D)=\theta_{S,R}-\theta_{T,R}$ for the Dirac $\nu_e$s (dashed  lines) and $\Delta \theta_{SP,R}=\theta_{S,R}-\theta_{P,R}(M)$ for the Majorana $\nu_e$s (dotted lines) when  $\theta_1=\pi/2$.
To explain the origin of the difference we give the formulas for  $A(\Phi)$ with assumed values of $\theta_1=\pi/2$, $\theta_\nu=\pi$, the experimental standard couplings and $E_\nu=1\,MeV$. For the Dirac scenario with  $V-A $, $S_R$ and  $T_R$ interactions, it reads:
\beq
\label{aziSRTRD}
\lefteqn{A_{D}^{(S,R)(T,R)}(\Phi)=
-\bigg\{\bigg[3.354(-\sin(\Phi)(8|c_S^R||c_T^R| \cos(\Delta\theta_{ST,R})}\nonumber\\
&& \mbox{} + 8 |c_T^R|^2 + 1.369)
  -⁠ 10|c_S^R||c_T^R| \sin(\Delta\theta_{ST,R}) \cos(\Phi) \\
 && \mbox{}  -⁠ 1.369 \sin(\Phi))    \bigg]/⁠
 \bigg[12(|c_S^R|^2 + 5 |c_T^R|^2 + 1.881) + 1.5( |2 c_S^R|^2 \nonumber\\
 && \mbox{} + 6 |c_T^R|^2 + 1.651)   +
8 (|c_S^R|^2  + 14 |c_T^R|^2 + 6.552)\nonumber\\
&& \mbox{} - 44|c_S^R||c_T^R|  \cos(\Delta\theta_{ST,R}) + 77.462    \bigg]\bigg\}. \nonumber
\eeq
The corresponding formula for  $A(\Phi)$ in the Majorana case with $V-A$, $S_R$, $P_R$ and $V+A$ couplings is of the form:  
\beq
\label{aziAllM}
\lefteqn{A_{M}^{(All)}(\Phi)=
    -⁠\bigg\{3.354\sin(\Phi)\bigg[ 4 |c_P^R| |c_S^R| \cos(\Delta\theta_{SP,R}) - 1.369\bigg] \nonumber} \\
&& \mbox{}   / \bigg[ 16( |c_P^R|^2 + 2.875|c_S^R|^2 + 4.841 )\bigg]\bigg\}. \nonumber
    \eeq
    It can be noticed that these expressions explicitly reveal different dependence on the azimuthal angle $\Phi$ in the two cases. Thus, it is not possible to reproduce the dashed curve of Fig. 3, which  describes the Dirac  $\nu$ beam, by using the all possible non-standard (flavour-conserving) Majorana $\nu$ interactions.\\
 Another possibilities are related to the $\mbox{\boldmath $\hat{\eta}_{\nu}$}\cdot\hat{\bf q}\not=-1$ case.
 When we assume that the incoming $\nu_e$ beam is the superposition of LC $\nu$s with RC ones and there is an experimental control of the angle $\phi_{\nu}$ connected with $\mbox{\boldmath $(\hat{\eta}_{\nu})^{\perp}$}$, we have new opportunities of testing the $\nu_e$ nature and TRSV. 
 Fig. 4 illustrates the asymmetry $A$ in this case. It probes the dependence of $A(\Phi_{max})$ (solid line) and the asymmetry axes location $\Phi_{max}$ (dashed line) on $\phi_{\nu}$. The plot compares behavior of the Dirac ($S_R$, $T_R$) and the Majorana ($S_R$) $\nu$s additionally taking into account TRSC (left column) and TRSV (right column) options. 
 The detection of a non--trivial dependence of the asymmetry on the $\phi_{\nu}$ shown in  Fig. 4 would indicate the existence of exotic scalar or tensor couplings of RC $\nu$s. The precise measurement of the magnitude of $A(\Phi_{max})$ would help to detect TRSV, however it can be difficult to distinguish between the Dirac and the Majorana $\nu_e$s.
 \\The Figs. (5-6) display the impact  of $\theta_1$ and $\phi_\nu$ on the possible values of the asymmetry $A_{\theta_e}(\Phi_{max})$ for the scalar interactions. We see that this measurement is sensitive to the presence of exotic couplings, offers the possibility of the distinction between the Dirac and the Majorana $\nu_e$s, and allows the detection of TRSV effects. For the sake of the illustration of the variety of possible outcomes we point on the most noticable characteristics of the diagrams presented in Fig. 5.
 The maximum values of $A_{\theta_e}(\Phi_{max})$  for $\theta_1=0.1$ increase  to $0.012$ in the Dirac case (dashed line in left upper plot), and to $0.035$  in the Majorana case (dashed line in right upper plot)  in  comparison  to the standard expectation of $0.004$, shown in Fig 2. When  $\theta_1=\pi/2$ the magnitude of $A_{\theta_e}(\Phi_{max})$ may decrease to around $0.02$ for the  Majorana $\nu_e$s (solid line in middle right plot), while the standard prediction gives $0.08$ at $\theta_e=\pi/6$. The  maximum value of $A_{\theta_e}(\Phi_{max})$ for $\theta_1=\pi-0.1$  decreases to $0.08-0.17$ for the Dirac $\nu_e$s (lower left plot) and to $0.2-0.27$ in the  Majorana case (lower right plot), compared to the standard expectation of $0.54$. In addition one can observe an offset of the maximum of $A_{\theta_e}(\Phi_{max})$ to higher values of $\theta_e$ particularly for the Majorana $\nu_e$s. Similar features can be observed in Fig. 6 for $\phi_{\nu}=\pi/4$. \\ 
It is necessary to  point out that from the experimental point of view searches of differences between the Dirac and the Majorana $\nu_e$s by the measurement of observables dependent on $\mbox{\boldmath $(\hat{\eta}_{\nu})^{\perp}$}$ would be extremely  difficult. In order to measure $A_{\theta_e}(\Phi_{max})$ one should determine the location  of $\Phi_{max}$ by counting the events from $\Phi$ to $\Phi+\pi$ and from $\Phi+\pi$ to $\Phi+ 2 \pi$ (for various  $\Phi$) at fixed $\theta_e$ (and a particular configuration of $\phi_{\nu}$). In this way $\Phi_{max}$ and $A_{\theta_e}(\Phi_{max})$ are found according to their definitions.
These measurements have to be  repeated for different $\theta_e$s. The experimental curve drawn with respect to  $\theta_e$ should fit to one of the curves in Figs.  2, 5, 6. The measurement of  $A(\Phi_{max})$ proceeds in a similar way, but now $\theta_e$ is not fixed - the azimuthal orientation of  $\mbox{\boldmath $(\hat{\eta}_{\nu})^{\perp}$}$  described by $\phi_{\nu}$ is fixed  instead. Events within the azimuthal angle from $\Phi$ to $\Phi+\pi$ and from $\Phi+\pi$ to $\Phi+ 2 \pi$ for all $\theta_e$ are counted.
The repetitions of the measurements for different  $\phi_\nu$s give a curve which should fit to one of the curves in Fig. 4.
\section{Distinguishing between Dirac and Majorana neutrinos via spectrum  and polar angle distribution of scattered electrons}
\label{sec5}
\begin{figure}
    \begin{center}
        \includegraphics[width=0.8\linewidth]
        {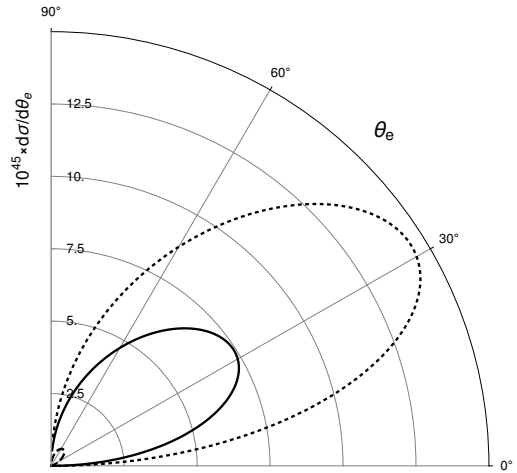}
        \caption{Dirac (or Majorana) $\nu_e$ with  $V-A$ interaction, $\mbox{\boldmath $\hat{\eta}_{\nu}$}\cdot\hat{\bf q}= - 1$, $E_\nu=1\,MeV$: plot  of $ d\sigma/d\theta_e$ as a function of $\theta_e$ for different values of $\theta_1$;
            dotted line for $\theta_1=0$; solid line for $\theta_1=\pi/2$; dashed line for $\theta_1=\pi$.
            \label{Fig.10}}
    \end{center}
\end{figure}
\begin{figure}
    \begin{center}
        \includegraphics[width=0.95\linewidth]
        {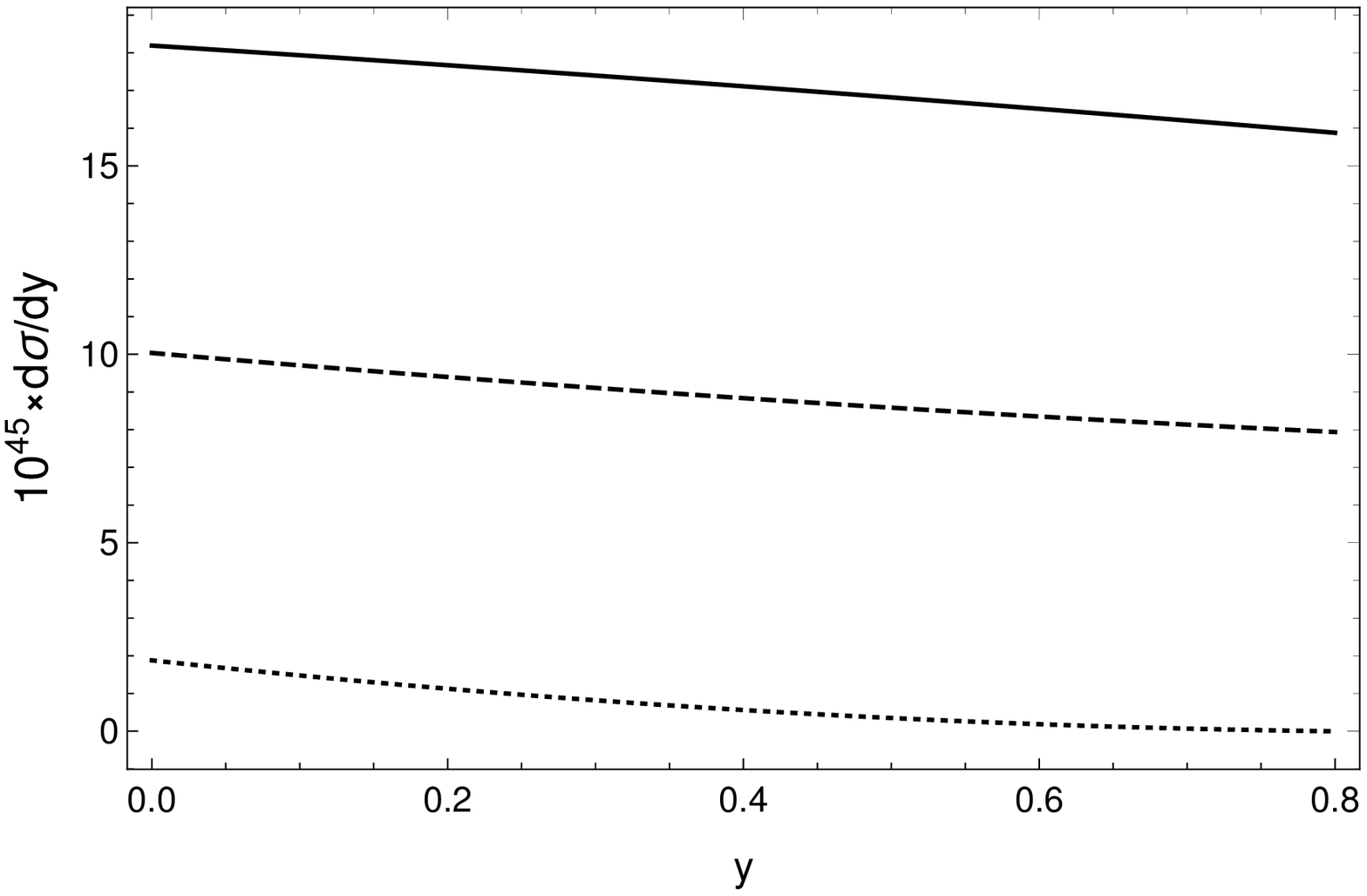}
        \caption{Dirac (or Majorana) $ \nu_e$ with  $V-A$ interaction  $\mbox{\boldmath $\hat{\eta}_{\nu}$}\cdot\hat{\bf q}= - 1$, $E_\nu=1\,MeV$:    plot  of $ d\sigma/dy$ as a function of $y$ for different values of $\theta_1$:
            solid line for $\theta_1=0$; dashed line for $\theta_1=\pi/2$; dotted line for $\theta_1=\pi$.
            \label{Fig.11}}
    \end{center}
\end{figure}
 \begin{figure}
 	\begin{center}
 		\includegraphics[width=0.75\linewidth]
 		{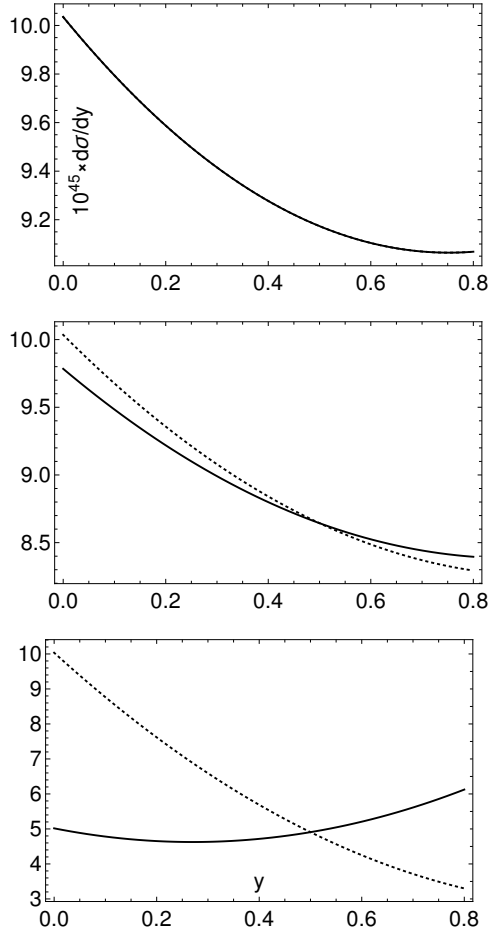}
 		\caption{$d\sigma/dy$ as a function of $y$ for  $V-A$ and $S_R$ scenario with $c_{S}^{R}=0.3$, $\theta_1=\pi/2$; $\mbox{\boldmath $\hat{\eta}_{\nu}$}\cdot\hat{\bf q}=-1$ (upper plot),
 			$\mbox{\boldmath $\hat{\eta}_{\nu}$}\cdot\hat{\bf q}=-0.95$ (middle plot), $\mbox{\boldmath $\hat{\eta}_{\nu}$}\cdot\hat{\bf q}=0$ (lower plot), for Dirac (solid line) and Majorana (dotted line) $\nu_e$s.
 			\label{y05}}
 	\end{center}
 \end{figure}
\begin{figure}
    \begin{center}
        \includegraphics[width=0.95\linewidth]
{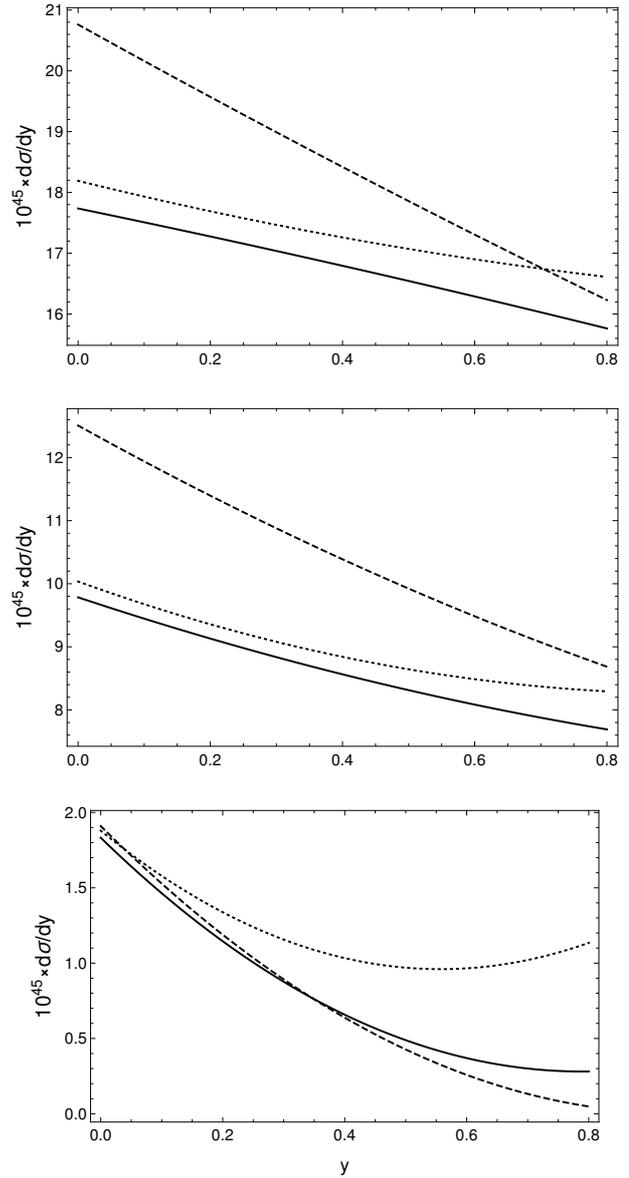}
        \caption{Superposition of LC $\nu_e$s with RC ones in presence of non--standard couplings with     $\mbox{\boldmath $\hat{\eta}_{\nu}$}\cdot\hat{\bf q}=-0.95$:    dependence of $d\sigma/d \,y$ on $y$  for different values of $\theta_1$ when $E_\nu=1\,MeV$,  $\phi_\nu=0$.
            Upper plot for $\theta_1=0$; middle plot for $\theta_1=\pi/2$; lower  plot for $\theta_1=\pi$;  dashed line for Dirac $\nu_e$ with  $V-A $ and $T_R$, $|c_T^R|=0.3, \theta_{T,R}=0$;  dotted line for Majorana $\nu_e$ with $V-A $ with $S_R$, $|c_S^R|=0.3, \theta_{S,R}=0$; solid line for Dirac  $\nu_e$ with $V-A $ and $S_R$ when $|c_S^R|=0.3, \theta_{S,R}=0$.\label{Fig.14}}
    \end{center}
\end{figure}
In this section we explore the $\nu_e$ nature problem by using the electron energy spectrum and the polar angle distribution of scattered electrons. To begin with, it is worth recalling that  the above observables do not allow one to differentiate between the Dirac and the  Majorana $\nu_e$s in the case of the  standard $V-A$ interaction in the relativistic limit; see Figs. (7-8) plotted for $\theta_1=0, \pi/2, \pi$.\\ If one assumes that the  $\nu_e$ source produces the superposition of LC with  RC $\nu$s, the cross sections $ d\sigma/d\theta_e$, $ d\sigma/dy$ for the detection of  Dirac and Majorana $\nu_e$s contain the interferences between LC and RC $\nu$s proportional to the various  angular correlations (T-even and T-odd) among  $\mbox{\boldmath $(\hat{\eta}_{\nu})^{\perp}$}$,  $\hat{\bf q}$, $\hat{\bf p}_{e}$, $\mbox{\boldmath $(\hat{\eta}_{e})^{\perp}$}$ vectors. Consequently the linear contributions from the non--standard interactions allow us to distinguish between the  Dirac and the Majorana $\nu_e$s, and  search for TRSV, see Figs. (9-10). \\
Fig. \ref{y05} shows that it is possible to distinguish the neutrino nature in the case of $V-A$ and $ S$ interactions. 
 Scalar couplings for both, the Dirac and the Majorana $\nu$s, are constrained to give the same value of the spectrum at some specified value of y 
 (in Fig. 9 $y=0.5$ is fixed). It follows that the remaining freedom cannot be used to adjust the two cross sections.
 The diagrams in Fig.9 illustrate the generic behavior of spectra although
 they are plotted for a particular value of $|c_S^R|=0.3$ defined for the Majorana $\nu_e$ (which, under the above constraint, determines a value of $|c_S^R|$ for the Dirac $\nu_e$). The transversal polarization of the incoming $\nu_e$
 plays here a crucial role,  because only for  $\mbox{\boldmath $(\hat{\eta}_{\nu})^{\perp}$}\not=0$  one can discriminate between  the Dirac and the Majorana neutrinos.
\\
Fig. 10 shows how the change of $\theta_1$ affects the  energy spectrum of recoil electrons in the  presence of interferences related to transversal component of the neutrino spin polarization, both for the Dirac and the Majorana $\nu_e$s; the plot should be compared with Fig.8. Small deviations for the low energy recoil electrons in the case of  Dirac scenario with  $V-A$ and $T$ interactions when  $\theta_1=0,\pi/2$ are seen (dashed line in upper and middle plots). On the other hand the departure from the standard prediction when $\theta_1=\pi$ is too small to be clearly visible (lower plot). 
It is worth noting that Figs. (9-10) have been made at fixed azimuthal angle $\phi_\nu=0$; changes of $\phi_\nu$ would affect the spectrum and polar distributions. 
\section{Conclusions} 
\label{concl}
We have studied the two theoretically possible scenarios of the physics beyond  the standard model in which flavour-conserving standard and non--standard interactions of both left chiral and  right chiral electron neutrinos were introduced. 
\\
We have shown that the various types of the azimuthal asymmetries of the recoil electrons, the energy spectrum and the polar angle distribution of the scattered electrons can in principle discern between the Dirac and the Majorana $\nu_e$s interacting with PET both for the longitudinal and the transversal $\nu_e$ polarizations.
The high-precision measurements of these quantities may shed some light on the fundamental problems of the $\nu$ nature and TRSV in the leptonic processes. 
In the particular case of the $V-A$ and $S$ couplings the spectrum can distinguish the two types of the neutrinos as long as the incoming $\nu_e$ beam has non--vanishing transversal polarization.
But even for the longitudinally polarized $\nu_e$ beam the asymmetry observable $A(\Phi)$ can identify the Dirac scenario with $V-A$, $S$ and $T$ interactions independently of any assumed Majorana non--standard couplings.
\\
The proposed new tests  require intensive monochromatic low-energy $\nu_e$ sources, large PET, and detectors enabling a measurement of  the azimuthal angle and the polar angle of the recoil electrons  with the  high angular resolution. Propositions of the relevant detectors have been discussed in the literature \cite{Hellaz,Hellaz1,Hellaz2,Heron,Heron1}. In turn high-resolution measurements of the  spectrum of low energy outgoing electrons demand
detectors with the ultra low detection threshold and background noise. Some interesting concepts of various (monochromatic) $\nu_e$  sources are under debate \cite{Source1,Source2,Source3,Source4,Source5,Source6,Source7}. A preliminary study of the feasibility of the electron polarized scintillating GSO target has been carried out by \cite{INFN}.
In order to make the detection of  $\mbox{\boldmath $(\hat{\eta}_{\nu})^{\perp}$}$-dependent effects possible, further studies on the appropriate choice of  $\nu_e$ source which would take into account the exotic couplings of RC $\nu_e$s are needed. This is necessary to explain the basic role of production processes in generating  $\nu_e$ beam with non--zero transversal polarization and to control the azimuthal angle $\phi_\nu$. The controlled production of $\nu_e$ beam with the fixed direction of $\mbox{\boldmath $(\hat{\eta}_{\nu})^{\perp}$}$ with respect to the production plane is impossible to date, thus the alternative option with the (un)polarized $\nu_e$ source generating only the longitudinally polarized $\nu_e$s seems to be presently more available.
\section{Appendix 1 - General formula on laboratory differential cross section for elastic  scattering of Majorana $\nu_{e}$s on PET}
\label{app1}
The laboratory differential cross section for Majorana $\nu_{e}$s, when
$\mbox{\boldmath $\hat{\eta}_{e}$} \perp {\bf \hat{ q}}$ ($\theta_1=\pi/2$),  is of the form:
\beq \label{przekMnue}
\lefteqn{ \frac{d^{2} \sigma}{d y d \phi_{e}} = \bigg(\frac{d^{2} \sigma}{d y d \phi_{e}}\bigg)_{V- A} + 
\bigg(\frac{d^{2} \sigma}{d y d \phi_{e}}\bigg)_{(S,P)_{R}} }\\
&&\mbox{} + \bigg( \frac{d^{2} \sigma}{d y d \phi_{e}} \bigg)_{V-⁠A}^{S_{R}}
 + \bigg( \frac{d^{2} \sigma}{d y d \phi_{e}} \bigg)_{V-A}^{P_{R}\nonumber}
\eeq
\beq \lefteqn{ \bigg( \frac{d^{2} \sigma}{d y d \phi_{e}} \bigg)_{V-⁠ A} =
    B \Bigg\{ c_{A}^2 \bigg[ \netpet \mbox{\boldmath $\hat{\eta}_{\nu}$}\cdot\hat{\bf q}(y-2)}\\
&& \mbox{} \cdot  \sqrt{y \left(\frac{2m_e}{E_\nu}+y\right)}
 +  y^2-⁠2y + 2 + \frac{m_e}{E_\nu} y    \bigg]    \nonumber\\
&& \mbox{} +
c_{V}^2 \bigg[ \netpet \mbox{\boldmath $\hat{\eta}_{\nu}$}\cdot\hat{\bf q}\; y\,\sqrt{y \left(\frac{2m_e}{E_\nu}+y\right)} \nonumber \\
&& \mbox{} + y^2-⁠2y +2 -⁠ \frac{m_e}{E_\nu} y    \bigg] \nonumber\\
&& \mbox{} + 2c_{V}c_{A}\bigg[\netpet\sqrt{y \left(\frac{2m_e}{E_\nu}+y\right)}(y-1) \nonumber\\
&& \mbox{} +y(y-2) \mbox{\boldmath $\hat{\eta}_{\nu}$}\cdot\hat{\bf q}\bigg]\Bigg\},\nonumber
\eeq
\beq
\lefteqn{\bigg(\frac{d^{2} \sigma}{d y d \phi_{e}}\bigg)_{(S,P)_{R}} = \mbox{}
    B\bigg\{ y\left(y+2\frac{m_{e}}{E_{\nu}}\right)
    |c_{S}^{R}|^{2} + y^{2}|c_{P}^{R}|^{2} }   \\
&& \mbox{} -4 y \sqrt{y\bigg(\frac{2 m_{e}}{E_{\nu}}+y\bigg)}\netpet\mbox{\boldmath $\hat{\eta}_{\nu}$}\cdot\hat{\bf q}\nonumber\\
&& \mbox{}\cdot\left[ Re(c_{S}^{R})Re(c_{P}^{R}) + Im(c_{S}^{R})Im(c_{P}^{R})\right]
\bigg\},  \nonumber\eeq
\beq
\lefteqn{\label{VmASR} \bigg( \frac{d^{2} \sigma}{d y d \phi_{e}} \bigg)_{V-A}^{S_R} = \mbox{}
2    B \Bigg\{
2     c_{V}  \sqrt{y\bigg(\frac{2 m_{e}}{E_{\nu}}+y\bigg)} }\\
&& \mbox{}  \cdot\bigg(\nutqpet Im (c_{S}^{R})  + \nutpet Re(c_{S}^{R})\bigg) \nonumber\\
&& \mbox{}  -  c_{A} \bigg(\frac{E_{\nu}}{m_{e}}y + 2 \bigg)\Bigg[ \petnetnut Im(c_{S}^{R})\nonumber\\
&&\mbox{} \cdot \sqrt{y\bigg(\frac{2 m_{e}}{E_{\nu}}+y\bigg)}
 + y\Bigg(\netpet\bigg(\nutqpet \nonumber\\
&&\mbox{} \cdot Im(c_{S}^{R})   +\nutpet Re(c_{S}^{R})\bigg)-Im(c_{S}^{R}) \nonumber\\
&&\mbox{} \cdot \qnetnut
 +  \frac{ m_{e}}{E_{\nu}}\netnut Re(c_{S}^{R})\Bigg)\Bigg]\Bigg\} \nonumber,
 \eeq
\beq
\lefteqn{\label{VmAPR} \bigg( \frac{d^{2} \sigma}{d y d \phi_{e}} \bigg)_{V-A}^{P_R} = \mbox{}
2    B \, c_{A}\,y\Bigg\{
      \frac{E_{\nu}}{ m_{e}}y\nutpet\bigg(
     Im (c_{P}^{R})}\\
    && \mbox{} \cdot \netqpet
    + \netpet  Re(c_{P}^{R})\bigg) \nonumber\\
&& \mbox{} + 2 \nutpet\bigg(\netqpet Im (c_{P}^{R}) +Re(c_{P}^{R})\nonumber\\
&& \mbox{} \cdot \netpet \bigg)
+ (y-⁠2)\netnut  Re(c_{P}^{R}))\nonumber\\
&& \mbox{} -  \sqrt{y\bigg(\frac{2 m_{e}}{E_{\nu}}+y\bigg)}\petnetnut Im (c_{P}^{R})  \nonumber
\Bigg\}, 
\eeq
where $B\equiv \left(E_{\nu}m_{e}/2\pi^2\right) \left(G_{F}^{2}/2\right)$.  $\mbox{\boldmath $\hat{\eta}_{\nu}$}$ is the unit 3-vector of
$\nu_{e}$ spin  polarization in its rest frame, Fig. 1. $(\mbox{\boldmath$\hat{\eta}_{\nu}$}\cdot\hat{\bf q}){\bf\hat{q}}$ is the longitudinal component of $\nu_{e}$ spin polarization.
$|\mbox{\boldmath $\hat{\eta}_{\nu}$}\cdot\hat{\bf q}| = |1-⁠ 2
Q_{L}^{\nu}|$, where $Q_{L}^{\nu}$ is the probability of producing the LC $\nu_{e}$.
We see that the interference terms  between standard $V-A$ and exotic $S_R, P_R$ couplings depend on  the transversal
$\nu_e$ spin polarization  $\mbox{\boldmath $(\hat{\eta}_{\nu})^{\perp}$}$ related to the production process (similar regularity  as in the Dirac case \cite{SBPET}).

\end{document}